\documentclass[12pt,preprint]{aastex}
\usepackage{amsmath}
\usepackage{accents}
\usepackage{graphicx}
\usepackage{natbib}
\usepackage{wrapfig}
\usepackage{subfigure}

\shorttitle{Imaging Exomoons}
\shortauthors{Peters \& Turner}

\begin{document}

\title{On the Direct Imaging of Tidally Heated Exomoons}

\author{Mary Anne Peters\altaffilmark{1,2}* and Edwin L. Turner\altaffilmark{1,3} }
\affil{$^1$ Department of Astrophysical Sciences, Princeton University, Princeton, NJ 08544, USA}
\affil{$^2$ Department of Mechanical and Aerospace Engineering, Princeton University, Princeton, NJ 08544, USA} 
\affil{$^3$ The Kavli Institute for the Physics and Mathematics of the Universe, The University of Tokyo, Kashiwa 227-8568, Japan}
\email{*correspondence: mapeters@princeton.edu}

\begin{abstract}
We demonstrate the ability of existing and planned telescopes, on the ground and in space, to directly image tidally heated exomoons orbiting gas-giant exoplanets. Tidally heated exomoons can plausibly be far more luminous than their host exoplanet and as much as $0.1\%$ as bright as the system's stellar primary if it is a low mass star. Because emission from exomoons can be powered by tidal forces, they can shine brightly at arbitrarily large separations from the system's stellar primary with temperatures of several hundreds degrees Kelvin or even higher in extreme cases. Furthermore, these high temperatures can occur in systems that are billions of years old.  Tidally heated exomoons may thus be far easier targets for direct imaging studies than giant exoplanets which must be {\it both} young and at a large projected separation (typically at least tens of AU) from their primary to be accessible to current generation direct imaging studies. For example, the (warm) Spitzer Space Telescope and the next generation of ground based instruments could detect an exomoon roughly the size of the Earth at a temperature $\approx600K$ and a distance $\approx5$ parsecs in the K-, L-, and M-bands at the 5$\sigma$ confidence level with a one hour exposure; in more favorable but still plausible cases, detection at distances of tens of parsecs is feasible. Future mid-infrared space telescopes, such as JWST and SPICA, will be capable of directly imaging tidally heated exomoons around the nearest two dozen stars with brightness temperature $\geq300K$ and $R\geq1R_\Earth$ orbiting at $\geq12AU$ from the primary star at a $5\sigma$ confidence level in a $10^4$ second integration. In addition it is possible that some of the exoplanets which have already been directly imaged are actually tidally heated exomoons or blends of such objects with hot young planets.  If such exomoons exist and are sufficiently common ({\it i.e.,} nearby), it may well be far easier to directly image an exomoon with surface conditions that allow the existence of liquid water than it will be to resolve an Earth-like planet in the classical Habitable Zone of its primary. \end{abstract}

\section{Introduction}
\label{sec:Introduction}
Direct imaging of exoplanets, especially those in the ``Habitable Zone" (HZ), is extremely difficult because of the high contrast ratio between the star and planet and because the of the very small star-planet angular separation. Indeed, all exoplanets that have been directly imaged to date are well separated from their host star, and are young systems that are still hot ($T_{eff}\sim1000K$) from their formation (as opposed to being heated by stellar irradiation).  Examples include the HR8799 planets, $\beta$ Pic b, LkCa15b and $\kappa$ And b (\citealt{Marois2008HR8799}, \citealt{Lagrange2008A-probable}, \citealt{Kraus2011LkCa} and \citealt{Carson2012Direct}). 

Although there has already been substantial discussion of the possibility of tidally heated exomoons (THEMs), extrasolar analogies of solar system objects such as Io, Europa and Enceladus, in the literature (\citealt{Peale1979Melting}, \citealt{Yoder1981The-tides}, \citealt{Ross1987Tidal}, \citealt{Ross1989Viscoelastic}, \citealt{Nimmo2007Shear}) and even of their potential astrobiological interest (\citealt{Scharf2006The-potential}, \citealt{Henning2009Tidally}, \citealt{Heller2013Exomoon}, \citealt{Heller2012Exomoon}), the possibility of detecting exomoons has so far been restricted to indirect methods (\citealt{Sartoretti1999On-the-detection}, \citealt{Han2002On-the-Feasibility}, \citealt{Simon2007Determination}, \citealt{Kipping2009aTransit}, \citealt{Kipping2009bTransit}).

This paper investigates the possibility that THEMs could be directly imaged (and perhaps already have been) with existing ground and space based instrumentation  and even more effectively with currently planned direct imaging facilities.  This scenario has several powerful advantages from an observational point of view.  THEMs may remain hot and luminous for periods of order a stellar main sequence lifetime and so could be visible around old stars as well as young ones. In addition, since THEMs may be hot even if they receive negligible stellar irradiation, they may be luminous at large separations from the system primary, thus reducing or eliminating the requirement of high contrast imaging capabilities.  Moreover, tidal heating depends so strongly on the orbital and physical parameters of an exomoon, that quite plausible systems ({\it i.e.,} with properties not very different from those occurring in the solar system) will result in terrestrial planet sized objects with effective temperatures as high as 1000K, or even higher in extreme but physically permissible cases.
 
In order to provide context and motivation for the analysis to follow, it is helpful to consider solar system tidally heated moons.  Io emits more energy per unit area at $\lambda \sim 5 \mu m$ than expected (\citealt{Witteborn1979Io}, \citealt{Spencer2005Mid-Infrared}) and has the highest measured temperatures of any body in the outer solar system due to tidal heating (\citealt{McEwen1997High-temperture}). If the Galilean moon system orbited Neptune with the semi-major axes of their orbits scaled down in proportion to the Roche radius of that planet (relative to Jupiter's), the bolometric luminosity of Io would be greater than that of Neptune. If a super-Io orbited Jupiter at it's current location, but was as massive and dense as Earth, it would be the brightest solar system object beyond 5 AU, out-shining even Jupiter in the  2-4$\mu m$ and 5.5-6.5$\mu m$ wavelength ranges\footnote{Based on \cite{Spiegel2012Spectral} model of a 1M$_J$ 1Gyr old cloud-free, solar metalicity Jupiter-like planet, and assumes a blackbody curve for the scaled up version of Io.}. Furthermore, if Io were as massive as Earth, it would be bright enough for JWST to detect at a distance of 5 parsecs!

The remainder of this paper is organized as follows:  We establish the exomoon tidal heating equations and present associated scaling relations in section \ref{sec:Equations}. In section \ref{sec:Observability} we discuss discovery space constraints and determine the detection limits and observability of THEMs with existing and future instrumentation.  Conclusions and implications are described in section \ref{sec:Conclusions}.

\section{Tidal Heating}
\label{sec:Equations}
Tidal heating of moons in the solar system, such as Io and Europa, have been analyzed in detail by \cite{Reynolds1987Europa}, \cite{Segatz1988Tidal} and \cite{Peale1978Contribution}. In this section we adapt the resulting equations for tidal heating of exomoons from these literature analyses. From these inputs the relevant scaling relations based on orbital, exomoon, exoplanet and host-star parameters are easily obtained.

\begin{figure*}
\centering
\includegraphics[scale = 0.78]{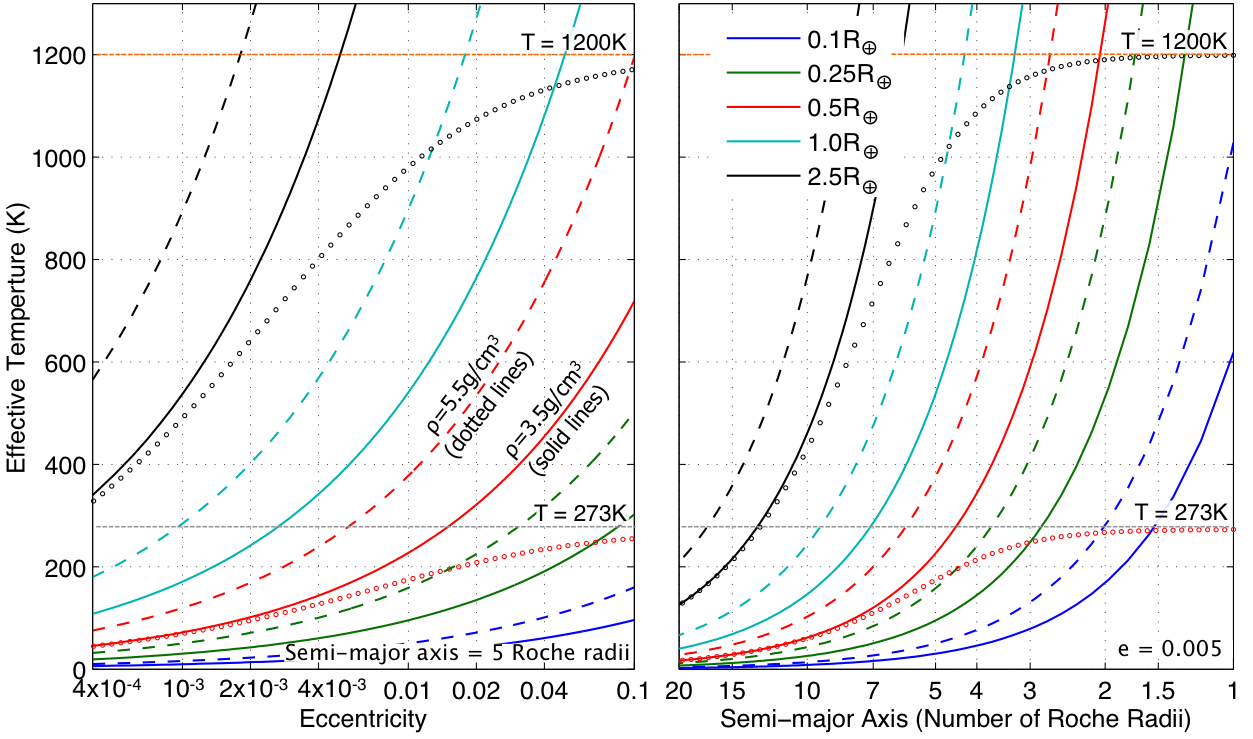}
\caption{LEFT: Plot of eccentricity vs. effective temperature of an exomoon, with a semi-major axis of 5 Roche radii (for comparison, Io is at 6.6 Roche radii). The solid and dashed lines correspond to a density of $\rho=3.5g/cm^3$ and $\rho=5.5g/cm^3$ respectively. The lower density ($\rho=3.5g/cm^3$) matches that of Io. The higher density ($\rho=5.5g/cm^3$) matches Earth. Each line color corresponds to a different moon radius listed in the upper left corner of the plot on the right (note that this legend lists the moon radii for both plots). The gray and orange dashed horizontal lines show the correspond to the melting temperature of water and rocks, respectively (note that rocks typically undergo a phase change between 900-1400K). The dotted red and black lines that approach the horizontal lines are meant to show that $Q$ and $\mu$ increase with increased tidal heating, and perhaps cause the THEM's temperature to plateau at some point which could plausibly coincide with a phase change. This is discussed in more detail in section \ref{sec:EffTempSED}. RIGHT: Plot of semi-major orbital axis vs. effective temperature of an exomoon. This plot assumes a moon with eccentricity of 0.005 (for comparison, Io has an eccentricity of $e = 0.004$). Dashed and solid lines represent the same densities as the plots on the left.}
\label{power laws figure}
\end{figure*}

\subsection{Luminosities}
\cite{Reynolds1987Europa} and \cite{Segatz1988Tidal} show that the average total luminosity of a moon due to tidal heating, $L_{tidal}$ is given by 
\begin{equation} \label{eq:Etidal}
	L_{tidal} = \frac{42\pi G^{5/2}}{19}\left(\frac{R_s^{7}\rho^{2}M_p^{5/2}}{\mu Q}\right)\left(\frac{e^2}{a^{15/2}}\right)
\end{equation}

where $G$ is the gravitational constant, $\mu$ is the moon's elastic rigidity, $Q$ is  the moon's dissipation function (or quality factor), $e$ is the eccentricity of the moon's orbit, $a$ is the semi-major axis of the moon's orbit, $\rho$ is the density of the moon, $R_s$ is the radius of the moon, and $M_p$ is the mass of the planet it orbits. Eq. \ref{eq:Etidal} assumes zero obliquity.

It is useful to eliminate the explicit dependence on the planet's mass by parameterizing Eq. \ref{eq:Etidal} in terms of the Roche radius. Let the moon's semi-major axis be some multiple, $\beta$, of the Roche radius, $a_{R}$, such that

\begin{equation} \label{eq:Roche}
	a = \beta a_{R} = \beta\left(\frac{3M_p}{2\pi\rho}\right)^{1/3}.
\end{equation}

Then we can rewrite the tidal energy flux equation as

\begin{equation} \label{eq:EtidalRoche}
	L_{tidal} = \left(\frac{6272\pi^7G^5}{9747}\right)^{1/2}\left(\frac{R_s^7\rho^{9/2}}{\mu Q}\right)\left(\frac{e^2}{\beta^{15/2}}\right)
\end{equation}

where we have grouped the terms that depend on the moon's physical properties and those that describe its orbit separately. Note that although $\beta$ is grouped with the orbital terms, in addition to its linear dependence on the moon's semi-major axis, $\beta$ is also more weakly dependent on the planet's mass and the moon's density.

Alternatively, we can write this equation in terms of $R_s$ and $M_s$ or $\rho$ and $M_s$ rather than $R_s$ and $\rho$. Alternative forms of Eq. \ref{eq:EtidalRoche} are
\begin{equation}
	L_{tidal} = \left(\frac{189G^{5/2}}{608\sqrt{2}\pi}\right)\left(\frac{M_s^{9/2}}{\mu Q R_s^{13/2}}\right)\left(\frac{e^2}{\beta^{15/2}}\right)
\end{equation}

\begin{equation}
	L_{tidal} = \frac{7}{19}\left(\frac{243\pi^7 G^{15}}{128}\right)^{1/6}\left(\frac{M_s^{7/3}\rho^{13/6}}{\mu Q}\right)\left(\frac{e^2}{\beta^{15/2}}\right)
\end{equation}

The scaling relation for Eq. \ref{eq:EtidalRoche} relative to the luminosity of Earth (which is $L_\Earth=$ 1.75$\times10^{24}$ ergs/s) is

\begin{equation} \label{eq:scaleTempLS}
	L_s  \approx L_\Earth \left[\left(\frac{R_s}{R_\Earth}\right)^{7}\left(\frac{\rho }{\rho_\Earth}\right)^{\frac{9}{2}}\left(\frac{36}{Q}\cdot \frac{10^{11} \frac{dynes}{cm^2}}{\mu}\right)\right] \left[\left(\frac{e}{0.0028}\right)^{2}\left(\frac{\beta}{8}\right)^{-\frac{15}{2}}\right].
\end{equation}

Note that Eq. \ref{eq:scaleTempLS} adopts $Q = 36$, $\mu = 10^{11}$ dynes/cm$^2$ and Earth's radius and density as reference values. The reference values of $\beta$ and $e$ were then chosen to give $L_{\Earth}$. Note that the first set of bracket terms in Eq. \ref{eq:scaleTempLS} includes exomoon's physical parameters, and the second contains the orbital parameters. This scheme is also used in Eqs. \ref{eq:scaleTemp} and \ref{eq:Wien} below.

\subsection{Effectives Temperatures and SEDs}
\label{sec:EffTempSED}
Using conventions of stellar astrophysics, we can define the exomoon's effective temperature, $T_s$, from the luminosity via the Stefan-Bolzmann Law
\begin{equation} \label{eq:Stefan}
	T_s  = \left(\left(\frac{392\pi^5G^5}{9747\sigma^2}\right)^{1/2}\left(\frac{R_s^5\rho^{9/2}}{\mu Q}\right)\left(\frac{e^2}{\beta^{15/2}}\right)\right)^{1/4} 
\end{equation}

where $\sigma$ is the Stefan-Bolzmann constant. Eq. \ref{eq:Stefan} can also be written as a scaling relation relative to a 279K (the equilibrium temperature of Earth) exomoon.

\begin{equation} \label{eq:scaleTemp}
	T_s  \approx  279K \left[\left(\frac{R_s}{R_\Earth}\right)^{\frac{5}{4}}\left(\frac{\rho }{\rho_\Earth}\right)^{\frac{9}{8}}\left(\frac{36}{Q}\cdot \frac{10^{11} \frac{dynes}{cm^2}}{\mu}\right)^{1/4}\right] \left[\left(\frac{e}{0.0028}\right)^{\frac{1}{2}}\left(\frac{\beta}{8}\right)^{-\frac{15}{8}}\right]
\end{equation}

This would give $\sim60K$ for Io, which would be the effective temperature of Io given no solar flux. Note that Eq. \ref{eq:scaleTemp} adopts the same reference values for $\mu$, $Q$, $\rho_s$, $R_s$, $\beta$ and $e$ as Eq. \ref{eq:scaleTempLS} since the reference temperature, 279K, corresponds to a blackbody the size of earth with the luminosity given by the reference luminosity in Eq. \ref{eq:scaleTempLS}. These are the $Q$ and $\mu$ values for Io from the literature (\citealt{Segatz1988Tidal}, \citealt{Peale1979Melting}). Eq. \ref{eq:scaleTemp} also assumes there is only tidal heating with no additional energy sources, such as stellar irradiation or interior radiogenic heat.  This is a conservative assumption since additional heat sources only serve to make the exomoon more luminous and thus easier to detect. We maintain that neglecting the stellar irradiation term is a good approximation for cases of substantial tidal heating and at THEM-star separations ($\gtrsim12AU$, and typically tens of AU) that are currently accessible to high-contrast instrumentation. For example, we can consider one of the most challenging direct imaging cases, an Earth-sized THEM heated to 300K and at a 12AU separation from its host star. For this case, the additional heating due to stellar irradiation is $<1\%$ of the tidal heating. However, should one want to calculate the temperature due to both tidal heating and stellar irradiation, it is given by

\begin{equation} \label{eq:insol_tidal}
	T_s  =  \left[\frac{L_{tidal} + L_{insol}}{4 \pi \sigma R_s^2} \right]^{1/4}
\end{equation} 

The scaling relations in Eq. \ref{eq:scaleTemp} are illustrated in Fig. \ref{power laws figure}. In Fig. \ref{power laws figure}, we adopt the $Q$ and $\mu$ values for Io in the literature. However, although $Q$ and $\mu$ are taken to be a constant for most of the curves shown in the figure, they are clearly not constant, even for the small number of objects in our solar system. Table 1 lists $Q$ and $\mu$ for several moons in the solar system as well as the Earth. From Table 1 we see that $Q$ and $\mu$ vary by two to three orders of magnitude and generally increase as the effective temperature of the moon increases. If these parameters increase as the tidal heating increases then one might expect a THEM's effective temperature to saturate for sufficiently strong tidal heating. It is likely that $Q$ and $\mu$ will change particularly rapidly at phase transitions such as melting for the material that constitutes the bulk of a THEM.
Certainly $T_{melt}$ will be different for rocky verses icy moons and the THEM's temperature could plausibly equilibrate near these melting points where the $Q$ and $\mu$ values might change very quickly by many orders of magnitude. We apply a function of the form

\begin{equation} \label{approxFunc}
	T_s  = \left(\frac{1}{T_{s,0}^2}+\frac{1}{T_{melt}^2}\right)^{-1/2}
\end{equation} 

to two of the curves in both the right and left plots of Fig. \ref{power laws figure} as a ``toy model" illustration of how the effective temperature might plateau at the melting temperature. The plateaus in Fig. \ref{power laws figure} given by Eq. \ref{approxFunc} are meant only to show that in general $L_{tidal} = L_{tidal}\left(\mu(L_{tidal}),Q(L_{tidal}) \right)$ and $T_{s} = T_{s}\left(\mu(T_{s}),Q(T_{s}) \right)$ (i.e. the tidal heating depends on $Q$ and $\mu$ which themselves depend on the tidal heating). Two values of $T_{melt}$ are shown as horizontal dashed lines in the plot. The grey line is the melting temperature of water. The orange line is at 1200K which is representative of the melting temperature of rock (the melting temperature of igneous rock is 800-1800K, \cite{Emiliani2007Planet}). The dotted red and black lines correspond to the solid red and black lines, respectively, but with Eq. \ref{approxFunc} applied. In short, it is quite possible that the extreme THEM temperatures shown if Fig. \ref{power laws figure} exist for some value of $R_s$, $\rho_s$, $\beta$ and $e$, however, it is difficult to estimate exactly what those values will be since the $Q$ and $\mu$ dependence on tidal heating at these extreme temperatures is poorly understood and depends on the composition of the THEM. 

Fig. \ref{power laws figure} still provides an understanding of the steep temperature dependence on $R_s$, $\rho_s$, $\beta$ and $e$ (for instance, the temperature goes almost as the square of $\beta$). On the other hand, the temperature dependency on $Q$ and $\mu$ is rather mild and goes only as the 1/4th power. Thus, these extreme temperatures possibly exist for reasonable values of $R_s$, $\rho_s$, $\beta$ and $e$, but likely for different (perhaps by orders of magnitude) values of $Q$ and $\mu$ than were used for the computation in Fig. \ref{power laws figure}. 

Given $T_s$, the peak wavelength of a exomoon's spectral energy distribution (SED) would be

\begin{equation} \label{eq:Wien}
	\lambda_{max}  \approx 10.4 \mu m \left[\left(\frac{R_s}{R_\Earth}\right)^{-\frac{5}{4}}\left(\frac{\rho }{\rho_\Earth}\right)^{-\frac{9}{8}}\left(\frac{36}{Q}\cdot \frac{10^{11} \frac{dynes}{cm^2}}{\mu}\right)^{1/4}\right] \left[\left(\frac{e}{0.0028}\right)^{-\frac{1}{2}}\left(\frac{\beta}{8}\right)^{\frac{15}{8}}\right]
\end{equation}

if it was emitting as an ideal blackbody.  However, solar system objects with significant tidal heating typically do not have a uniform temperature surface emitting at each point as an ideal blackbody.  Rather they display ``hot spots" and even vulcanism, locations on the surface through which a larger, often much larger, than average part of the internal heating is being radiated away.  This implies an SED that deviates significantly from a Planck form and which, in particular, emits more at shorter wavelengths than the uniform temperature blackbody approximation indicates.  This is seen in the SED of Io, for example (\citealt{Spencer2005Mid-Infrared}). However, modeling of the complexities of heat transport in the interior of a THEM far exceeds the scope of this initial discussion of detectability, and thus we hereinafter adopt the simple blackbody model of THEM SEDs.  In most respects, this is a conservative assumption in that it makes them less detectable than they would be expected to be with a more realistic SED. However, we note that this is not always true as the presence of hot spots will shift the SED to bluer wavelengths where the THEM will have to compete against an increased amount of star light. The blackbody SED assumption will be more accurate for an exomoon on which a thick atmosphere and/or oceans effectively redistribute the tidal heat emitted at hot spots on its surface. 

\subsection{Contrast}
Discussions of direct imaging of exoplanets are typically heavily focused on issues of contrast and associated instrumental inner working angles for the star-exoplanet separation on the plane of the sky.  The associated considerations for THEMs are more complex since both contrast with the planet orbited by the exomoon and contrast with the stellar primary must be taken into account.  Since the star-exoplanet separation does not affect the tidal luminosity of a given exomoon system at separations large enough to permit direct imaging, the contrast with the star may not (or may) be an important observational issue.  However, since tidal heating is very sensitive to the moon-planet separation, it is not plausible that a significantly bright THEM can be resolved from the planet it orbits with existing or planned facilities.  In other words, any emission from an exoplanet will dilute that from any THEM which orbits it. Furthermore, one can imagine a scenario where the THEM emits more light at certain wavelengths than the exoplanet and visa-versa. Note the fact that current instrumentation is only capable of directly imaging THEMs at large separations which allow us to ignore nuances such as perturbations from the star on the satellite's orbit which could induce variations on the exomoon's eccentricity and tidal heating ({\citealt{Cassidy2009Massive}).

Tidally heated moons are easier to detect if $L_s$ is large and $L_p$ and $L_*$ are small. Eq. \ref{eq:EtidalRoche} gives the tidal luminosity of an exomoon.  The contrast of the planet with the moon ($L_p/L_s$) decreases for colder exoplanets ($L_p \propto T_p^4$) that are further away from their host star and for older exoplanets which have already cooled from their initial formation temperatures. The planet-moon contrast also decreases for an exoplanet with less surface area ($L_p \propto R_p^2$). The contrast of the star relative to the moon ($L_*/L_s$) decreases for less massive stars ($L_* \propto M_p^{2.3}$ for $M<0.43M_\Sun$ and $L_* \propto M_p^{4.0}$ for $M>0.43M_\Sun$; \citealt{Burrows2001The-Theory} and \citealt{Duric2004Advanced}). Finally, the contrast requirement will be relaxed for relatively nearby star systems due to the resulting larger angular separations on the sky.

  \begin{table}\label{Table:SolSysMoons}
  \begin{center}
  \caption{Relevant physical parameters for various solar system bodies}
  \begin{tabular}{ l c c c c c c}
  \hline
   Parameter\tablenotemark{a} & Io & Europa & Enceladus & Moon & Titan & Earth \\ \hline \hline
   $a$ ($10^3$km) & 422 & 671 & 238 & 384 & 1222 & - \\ 
   $\beta$  & 6.63 & 10.0 & 3.82 & 40.5 & 23.3 & -  \\     
   $e$ & 0.0041 & 0.0101 & 0.0045 & 0.055 & 0.029 & - \\ 
   $R$ (km) & 1821 & 1560 & 250 & 1737 & 2575 & 6371 \\ 
   $\rho$ (g/cm$^3$) & 3.53 & 3.02 & 1.61 & 3.35 & 1.88 & 5.52\\ 
   $\mu$ ($10^9$ $\frac{dynes}{cm^2}$) & (100)\tablenotemark{b} & (40) & (2000) & 50 & (9) & 1200\\ 
   $Q$ & (36) & (100) & (100) & 27 & (100) & 280 \\  
     $T_{eff}$ (K) & 110 & 102 & 75 & 225 & 94 & 287 \\ \hline
   \tablenotetext{a}{References for values are \cite{Lodders1998The-Planetary}, \cite{Segatz1988Tidal}, \cite{Scharf2006The-potential}, \cite{Murray1999Solar}, \cite{Yoder1995Astrometric}, \cite{Thomson1863Dynamical}}
   \tablenotetext{b}{Parenthesis indicate this is a modeled, not measured value.}
   \end{tabular}
   \end{center}
   \end{table}

\subsection{Lifetime}
In order to sustain tidal heating, a moon must preserve its orbital eccentricity. In some cases, the eccentricity is maintained by resonance with another moon. For example, Io, Europa and Ganymede are in a 1:2:4 mean motion resonance that sustains the former's orbital eccentricity. Io is being heated by tidal forces 4.5 Gyr after the formation of the solar system, which suggests tidal heating can occur in old planetary systems. Hence, we are likely to find THEMs in systems where they are in orbital resonances with other exomoons; however, there is no reason to believe that such circumstances are uncommon. Both Jupiter and Saturn have close-in moons that participating in orbital resonances (\citealt{Peale1976Orbital}), and there are theoretical reasons to expect such resonances to develop naturally during the formation process  (\citealt{Yoder1979How-tidal}, \citealt{Cassidy2009Massive}, \citealt{Ogihara2012N-body}). Furthermore, it is also possible for an exomoon to maintain an eccentric orbit via perturbations from other planets (\citealt{Matija2007Excitation}) or by the star (\citealt{Georgakarakos2002Eccentricity}, \citealt{Georgakarakos2003Eccentricity}).

In general, more massive planets and more massive moons are expected to allow longer lifetimes for the orbital resonances that sustain the tidal heating at fixed tidal heating luminosity.

There is evidence for strong tidal dissipation in Io and Jupiter (\citealt{Lainey2009Strong}). To determine if such enormous amounts of tidal heating can be sustained over the lifetime of a planetary system, we can divide the orbital energy of the moon and rotational energy of its host planet by the THEM's luminosity. It turns out that the rotational energy of Jupiter is substantially larger than the orbital energy associated with Io, and the orbital energy term can be ignored for the Jovian system. 
A few percent of Jupiter's rotational energy could provide enough energy to maintain Io at $\approx300K$ for the age of the solar system. And for a larger planet or smaller THEM, temperatures of 1000K could be sustained for billions of years. Io's orbital energy alone (excluding Jupiter's rotational energy) could sustain a 300K Io for of order 100Myrs.

\subsection{Variability}
In general the observed brightness and SED of a THEM is expected to be substantially variable for multiple reasons:  Most dramatically the exomoon may be eclipsed by the much larger and darker exoplanet it orbits, thus causing a sharp drop (and later increase) in the observed flux.  The steep dependence of tidal heating on semi-major axis implies that the most luminous exomoons will have close in orbits and thus particularly large eclipse probabilities.  Even in the absence of eclipses, phase curve variations are expected since tidal heating typically produces moderate to extreme temperature variation across an object's surface.  Moreover, even at a single location on the exomoon's tidally heated surface, the temperature may well fluctuate due to time varying transport of interior heat to the surface, {\it e.g.,} vulcanism.  In general short timescale (hours to days would be expected from solar system analogs) variability would be a signature of THEMs which would help distinguish them from the relatively steady emission expected from a cooling gas-giant exoplanet. The period of the orbit, $T_{orbit}$ is just

\begin{equation} \label{eq:period}
	T_{orbit} = \frac{2\pi a^3/2}{\sqrt{GM_p}}.
\end{equation}

And the fraction, $F_{ellipse}$ of that time that the THEM spends in ellipse, $\tau_{ellipse}$, assuming a circular, coplanar orbit  (given in \citealt{Heller2012Exomoon}) is

\begin{equation} \label{eq:ellipse}
	F_{ellipse} = \frac{R_p}{\pi a} 
\end{equation}

where $R_p$ is the radius of the planet. For Io the parameters in Eq. \ref{eq:period} and \ref{eq:ellipse} are, $T_{orbit}$ = 1.77 days and $F_{ellipse}$ = 5.4\% respectively.

\subsection{Spectral Signatures}

Because THEMs can be much smaller than objects that are kept warm by their own internal heat (such a brown dwarf or a giant planet) and much hotter than they would be due to stellar irradiation from their parent stars, some THEMs are expected to be easily and conclusively identified by their photometric properties (given that the distance to the system is known, which is very likely to be the case). In particular, rocky THEMs would have an SED significantly too blue for their maximum stellar irradiation temperature and a flux much too small to be consistent with a giant planet or brown dwarf. For example, an Earth sized object with a surface temperature of 800 K seen at a projected (and thus minimum) separation of 30 AU from a G-dwarf star would have an apparent magnitude and colors quite inconsistent with either a cooling brown dwarf or giant planet.

 In this paper we have approximated THEM spectra as blackbodies. While this is perhaps a reasonable rough approximation, it is worth noting that deviations are expected from this simplistic model. The blackbody model gives us moons that will be hotter and smaller than their host exoplanets. One correction is that THEMs would likely have an excess emission at bluer wavelengths due to hot spots on their surfaces compared to that expected for a single effective temperature model (as is the case for Io, see section 2.2). Additionally, absorption features could substantially modify their SEDs, depending on the THEM's surface composition. THEM spectral absorption features might be similar to those observed in the SEDs of lava on Earth (\citealt{Oppenheimer1998Remote}, \citealt{Ramsey1999Estimating}) or Io (\citealt{McEwen1998High}, \citealt{Schmitt1994Identification}, \citealt{Geissler2003Volcanic}) or even similar to the models of extremely hot, rocky exoplants (\citealt{Hu2012Theoretical},  \citealt{Kaltenegger2010Detecting}). Unfortunately, fully predictive modeling of THEM SEDs would be complex and very probably underdetermined since it would require an understanding of the temperatures and distribution of hotspots as well as assumptions about surface and atmosphere composition, pressure etc. Nevertheless, SEDs indicating surprisingly high temperatures (given the stellar irradiation) and small (relative to giant planet sizes) luminosities are still likely to be valuable and relatively reliable indicators of the presence of a THEM.

We note in passing that the data collected on Fomalhaut b prior to the recent detection of F435W emission (\citealt{Currie2012Direct}) fits the general SED properties expected for a THEM. Specifically, aside from the F435W flux, Fomalhaut b is consistent with a THEM that is a modest fraction of the size of Io and has a surface temperature of $\sim$1600K if (1) an absorption feature in H-band is present and (2) surface hotspots lead to increased emission at bluer wavelengths ({\it e.g.} the F606W flux). However, the new F435W point in the object's SED curve would require the presence of unrealistically high temperature ($\sim$6000K) hotspots. Moreover, the optical colors of Fomalhaut b are a good match to those of its primary star, thus strongly suggesting a scattered stellar radiation explanation (\citealt{Currie2012Direct}). It therefore appears quite unlikely that Fomalhaut b is a THEM. Nevertheless, it remains an interesting example of an object with some of the spectral characteristics expected for a THEM.

\section{Detectability}
\label{sec:Observability}
\subsection{Temperature Limit}
\label{sec:Temperature}

Before exploring the THEM discovery space it is necessary to understand constraints on exomoon temperatures that can be produced by tidal heating. The temperature power law dependencies are so strong that the simple scaling relations presented in the previous section yield temperatures of thousands of degrees for seemingly plausible hypothetical exomoon systems. However, there are other physical constraints that limit the effective temperatures that can be achieved by tidal heating. These constraints are discussed in \cite{Cassidy2009Massive}. The \cite{Cassidy2009Massive} results suggest that mass loss from the satellite could erode THEMs on a Gyr timescale. We note that long-lived resonances in the solar system exist, and thus should exist in other systems as well.

The \cite{Cassidy2009Massive} analysis does not yield any precise upper limit on the surface temperature of an exomoon; we therefore only present calculations of THEMs up to  $T_s = 1000K$ in our following analysis. As context for this number note that the surface temperatures of known rocky bodies, such as Mercury and UCF-1.01 in the GJ 436 system, are of this order.  The Sun facing side of Mercury reaches temperature of 700K, and UCF-1.01's surface is estimated to be at $\sim860K$ (\citealt{Stevenson2012Two-nearby}). Given that rocky planets exist at surface temperatures near $1000K$ and that some rocks have even higher melting temperatures, it is plausible that rocky exomoons can survive at or above 1000K. 

\subsection{Assumed Exomoon Properties}
\label{sec:Assumptions}
   
The quantities $Q$ and $\mu$ are poorly known even for most solar system objects. For the first-approximation models considered in Fig. \ref{power laws figure}, we adopted the values for Io in the literature and then discussed where this assumption was valid. The following calculations for Fig. \ref{busy figure ground} and \ref{busy figure} do not require that we assume a particular value of $\mu$ or $Q$. They only require that we choose the radius of a moon, and assume that there is some combination of the remaining parameters in Eq. \ref{eq:scaleTemp} ({\it e.g.} $R_s$, $\rho_s$, $Q$, and $\mu$) that will give us the desired luminosity and hence temperature shown in the following plots. We will discuss the physical implications and viability of those hypothetical exomoon parameterizations.

\subsection{Existing Facilities}

\begin{figure*}
\begin{minipage}{\textwidth}
\centering
\includegraphics[scale = 0.72]{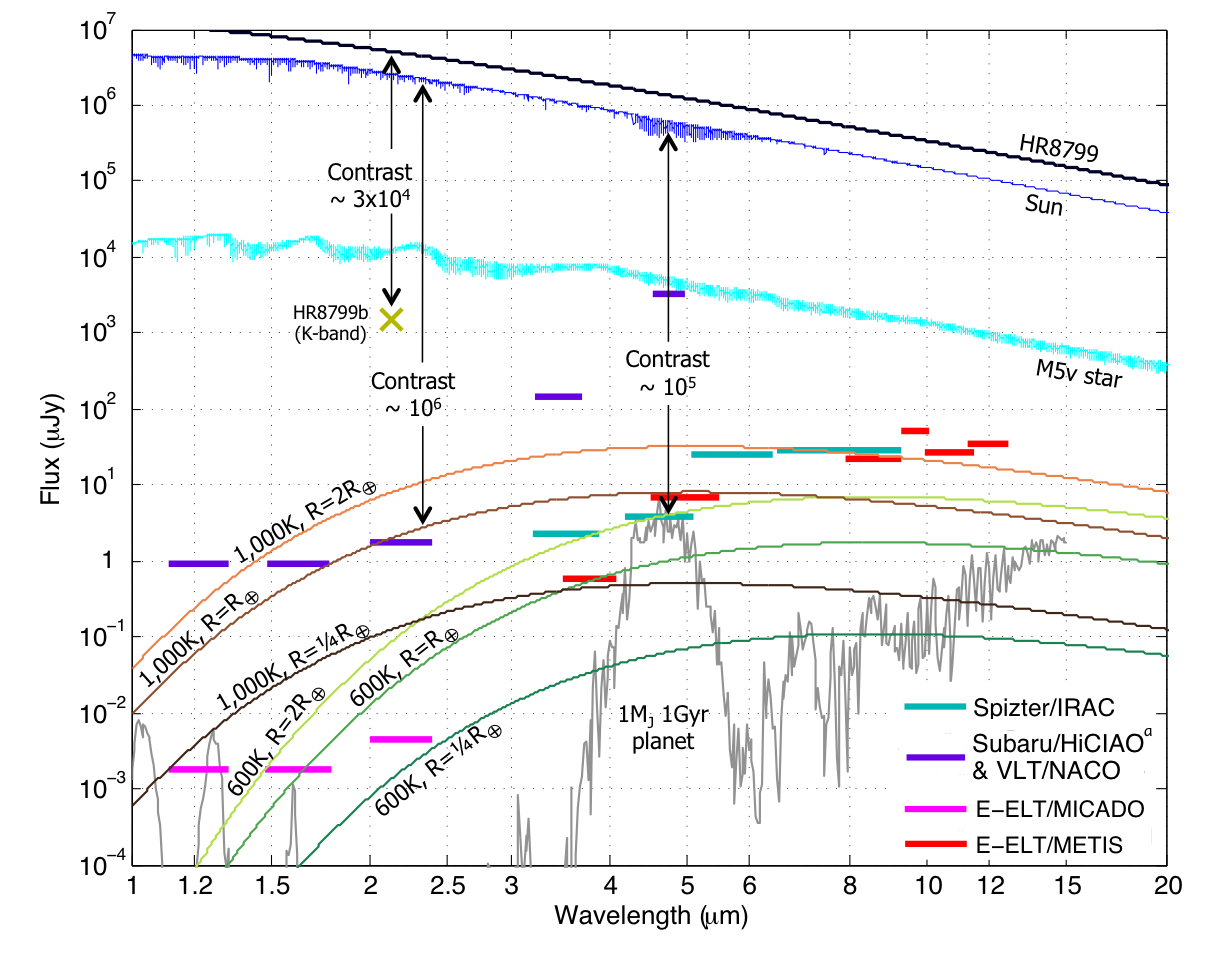}
\caption[Caption for LOF]%
{Plot shows the flux vs. wavelength of stars, an exoplanet, and various THEMs as well as the 1 hour integration time, 5$\sigma$ sensitivity of many instruments. The top three lines are the energy flux received from HR8799 (black line), the Sun (blue) and an M5V star (light blue) at a distance of 5 parsecs. The lines labeled with a temperature and radius are six plausible THEM blackbody curves. Note that Io is approximately one quarter the radius of Earth. The light gray line is a modeled exoplanet 1M$_J$ in mass, 1Gyr old and with no clouds and solar metallically (\citealt{Spiegel2012Spectral}). Note that this exoplanet is younger ({\it e.g.} brighter) than most stars in the solar neighborhood. The three vertical black lines show the contrast between a THEM and star for three different cases. The horizontal bars are the detection limits for Spitzer/IRAC (teal), Subaru/HiCIAO and VLT/NACO (purple)\footnote{{\footnotesize HiCIAO and NACO have equivalent sensitivities, however HiCIAO operates in J-, H-, and K-bands whereas NACO operates in these three bands as well as the L- and M-bands.}}, E-ELT/MICADO (magenta) and E-ELT/METIS (red).}
\label{busy figure ground}
\end{minipage}
\end{figure*}

Fig. \ref{busy figure ground} illustrates the detectability of tidally heated exomoons with existing ground and space-based facilities. The sensitivity of two future ground based instruments is also shown for comparison. The exomoon curves are shown as blackbodies at a given temperature and radius. The curves shown for the three stars are from the Kurucz models (\citealt{Kurucz1970Atlas}). Note that a 2R$_{\earth}$ moon would have a mass of 5-20M$_{\earth}$ whereas the expected theoretical upper limit is near 1M$_{\earth}$ (\citealt{Canup2006A-comman}, \citealt{Ogihara2012N-body}). Although the theoretical limits discussed in these two papers implies that the 2R$_{\earth}$ THEMs shown in Fig. \ref{busy figure ground} do not exist, it is possible that moons form via other processes not discussed in \cite{Canup2006A-comman} and \cite{Ogihara2012N-body}. For instance, the process that created the Earth's Moon produced a moon-to-planet mass ratio which is equivalent to a 20M$_{\earth}$ moon orbiting a 5M$_J$ exoplanet. The Moon's formation is difficult to model (\citealt{Canup2001Orgin}), and it is feasible that a analogous process occurs for gas giants that is capable of producing moons with $>$1M$_{\earth}$.

The instrument sensitivities are shown for a $5\sigma$, one hour integration detection limit. The VLT's NACO and Subaru's HiCIAO ground-based instruments are two of the most sensitive high-contrast imaging instruments currently on sky. The high contrast and adaptive optic (AO) systems currently on the infrared platform at the Subaru telescope, namely HiCIAO (\citealt{Hodapp2008HiCIAO}) and AO188 (\citealt{Hayano2010Commissioning}, \citealt{Minowa2010Performance}), are currently capable of achieving $10^{-5}$ contrast at 0.2 arcsec angular separation from the host star. The implementation of the next generation of high contrast instrumentation, SCExAO (\citealt{martinache2011subaru}) and CHARIS (\citealt{McElwain2012Scientific}) will be able to do at least an order of magnitude better than that ($10^{-6}$ contrast) at the same separation, and $10^{-7}$ contrast at a 2 arcsec separation. Based on Fig. \ref{busy figure ground}, the ability of these instruments to detect THEMs is likely to be limited by their sensitivity rather than their achievable contrast. It is possible future ground based instrumentation operating in J-, H-, and K-bands (such as MICADO, \citealt{Davies2010MICADO}) will be contrast, rather than sensitivity, limited unless observing late-type stars. However, the high contrast exoplanet imagers, such as E-ELT's EPICS (\citealt{Kasper2010EPICS}), claim they will be able to achieve contrasts of $2\times10^{-10}$ at 0.2 arcsec angular separations, which is again likely to make directly imaging THEMs a sensitivity and not a contrast problem.

The high contrast planet imagers (such as GPI, SPHERE and CHARIS) (\citealt{Macintosh2008gemini}, \citealt{Beuzit2008SPHERE}, \citealt{Peters2012Conceptual}) coming on sky in the next 1-3 years should be able to detect 800K, one Earth-radius moons in the K-band in systems with late-type M-dwarfs at a distance of a few parsecs. This same exomoon would have even more favorable contrast with it's host star at $\lambda = 4.5\mu m$ with warm Spitzer. It is interesting to note that at these temperatures, the exomoon would likely be many orders of magnitude brighter than its host planet at most wavelengths, even if the planet were $10\times$ more massive than Jupiter, assuming a system with an age comparable to the solar system's.

Beyond K-band, Warm Spitzer is currently the most sensitive telescope. Spitzer's IRAC could detect a 850K, $1R_\Earth$ moon at 5 parsecs. Fig. \ref{busy figure ground} shows that THEM would be less than $10^4$ times dimmer than a late-type M-dwarf at these wavelengths. The next generation of ground based instrumentation with comparable spectral coverage (such ars ELT's METIS, \citealt{Kendrew2010Mid}) will be even more sensitive than Spitzer and have a much higher angular resolution (see METIS detection limits in Fig. \ref{busy figure ground}). The ELT's METIS and MACADO should be able to detect 600K THEMs with radii of $0.5-1R_\Earth$.

\subsection{Future Facilities}

\begin{figure*}
\centering
\includegraphics[scale = 0.7]{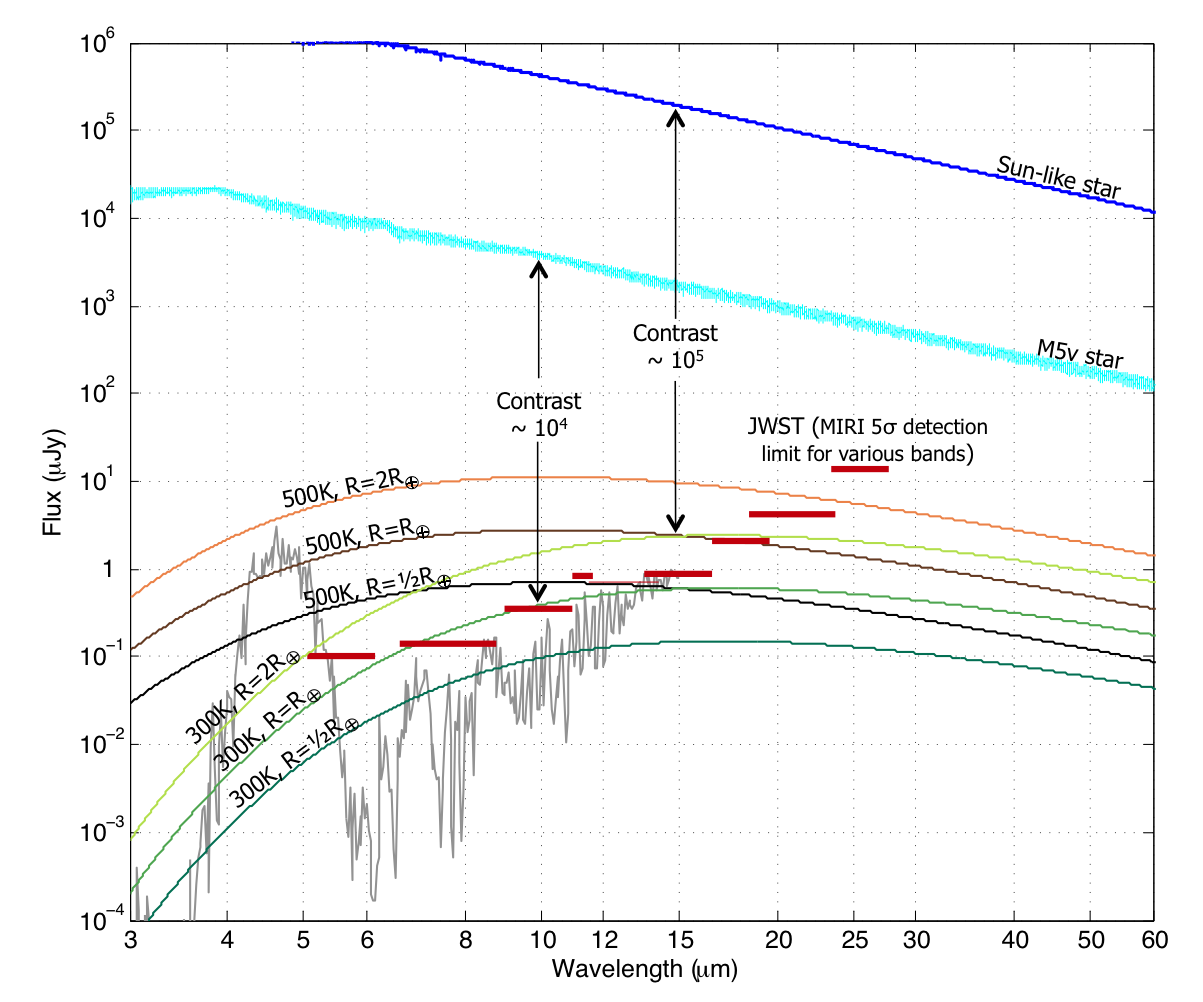}
\caption{The top two blue lines are the energy flux received from the Sun (blue) and M5V star (light blue) at a distance of 3 parsecs. The lines labeled with a temperature and radius are four plausible exomoon blackbody curves. The gray line is the 1Gyr, 1M$_J$ planet in Fig. \ref{busy figure ground}, but scaled to the luminosity of Jupiter so that it represents an older planet, more typical of the ages of planets found in the local neighborhood.  The red horizontal lines are the $10^{4}$ seconds integration time, 5$\sigma$ detection limit for JWST. Left contrast line shows that a 300K, Earth-radius moon would be $10^{4}\times$ fainter than the M5V star. The contrast line on the right shows that a 300K, $R = 2R_\Earth$ object would be $10^{5}\times$ fainter than a Sun-like and $\sim5\times$ brighter than a Jupiter. Note that the x- and y-axis are different than in Fig. \ref{busy figure ground}.}
\label{busy figure}
\end{figure*}

Future instruments, for example JWST's Mid-Infrared Instrument (MIRI), offer even more potential for direct imaging of exomoons. Fig. \ref{busy figure} shows the discovery space for MIRI. Note that this discovery space should be similar to SPICA's with a slight correction for SPICA's marginally smaller aperture. Similar to Fig. \ref{busy figure ground}, the solid lines labeled with temperatures and radii correspond to exomoon blackbody curves with those parameters. Note that the exomoons shown here have temperatures similar to Earth's rather than the much hotter temperatures shown in Fig. \ref{busy figure ground}. Thus, it is plausible that some of the exomoons JWST is capable of detecting, could potentially be habitable, in the sense of having surface temperatures that would allow liquid water to be present. Some of these exomoons have comparable irradiance to the gas giants in our solar system. At $\sim14\mu m$ a 300K, Earth-radius exomoon would be as luminous as Jupiter. However, if Jupiter were colder  due to being less heated by its primary and/or being older, the Earth-like moon would be much brighter than the planet.  The red bars are the $5\sigma$ detection limits for JWST-MIRI with a 10,000s integration time (\citealt{Glasse2010The-throughput}).  

A 300K, Earth-radius tidally-heated exomoon is only $3\times10^5$ times fainter at $\lambda\sim14\mu m$ than a Sun-like star but can be at a large distance from it's host star. For example, at 30AU projected separation, it would be 15 arcsecs from the star at a distance of 2 parsecs. At $\lambda=14\mu m$, $\lambda/D$=0.44 arcsecs for a 6.5m telescope, which means this moon would be at $30\lambda/D$. This far away from the star, the airy rings are $\sim3\times10^5$ times fainter than the core of the star and about the same intensity as the 300K, Earth-radius moon, indicating that the detection should be possible. This example is not at the limit of JWST's sensitivity and inner working angle. The most challenging THEM detection JWST will be capable of making is a 300K THEM as far as 4pc from the sun. If there is such an Earth-sized 300K moon orbiting $\alpha$Cen, MIRI will be able to detect it in 8 of its 9 spectral bands with better than 15 sigma signal-to-noise in a $10^4$ sec integration. Thus, directly imaging a 300K, Earth-radius moon that is tidally heated is potentially much easier than resolving an Earth-like exoplanet orbiting in the HZ of its primary. 

SPICA is another future space telescope that is ideal for exomoon detection. The SPICA coronagraph is being designed to operate from $3.5-27\mu m$ with a contrast of order $10^{-6}$ and a $3.3\lambda/D$ inner working angle (\citealt{Enya2011The-SPICA}). SPICA has a slightly higher noise floor than JWST, but should be able to achieve similar contrast. Thus SPICA's exomoon discovery space will be similar to JWST's, though JWST's larger aperture will give MIRI a modest advantage.

The next generation of ground based telescopes such as the Giant Magellan Telescope (GMT, \citealt{Johns2008The-Giant}), the Thirty Meter Telescope (TMT, \citealt{Nelson2008The-Status}), and the European Extremely Large telescope (E-ELT, \citealt{McPherson2012Recent}) are best suited for detecting THEMs at closer exomoon-star separations and at slightly shorter wavelengths than SPICA and JWST (and are therefore plotted in Fig. \ref{busy figure ground} which has shorter wavelengths plotted on the x-axis) and discussed in the previous section.

\section{Conclusions}
\label{sec:Conclusions}

Direct imaging detection of physically plausible, tidally heated exomoons is possible with existing telescopes and instrumentation.  If tidally heated exomoons are common, for example if typical gas giant exoplanets are orbited by satellite systems broadly similar to those found in the solar system ({\it i.e.} if Io was the same radius as Titan and at a similar number of roche radii as Enceladus -  roughly twice as close to Jupiter), we are likely to be able to image them around nearby Sun-like stars in the midst of their main sequence lifetimes with near future facilities.

Existing instrumentation should be able to detect exomoons with temperatures $\geq600K$ and $R\geq1R_\Earth$. Future mid-infrared space telescopes such as JWST and SPICA will be capable of directly imaging tidally heated exomoons around the nearest two dozen star systems with brightness temperature  $\geq300K$ and $R\geq1R_\Earth$ orbiting at $\geq12AU$ around stars within 4 parsecs of Earth at the $5\sigma$ confidence level in multiple bands. It is possible that some of the exoplanets which have already been directly imaged could be THEMs or exoplanet-THEM blends. It is therefore plausible that a habitable (in the sense of possessing liquid water on its surface) exomoon can be imaged long before it will be possible to do so for a habitable (in the same sense) exoplanet heated primarily by stellar irradiation.

Thus, the era of astrobiology based on direct imaging of extrasolar objects may not have to await the advent of specialized space-based telescopes such as those contemplated for the TPF and DARWIN missions.

\section*{Acknowledgements}
We thank Alexis Carlotti, Ren{\'e}  Heller, Jill Knapp, Matt Mountain, George Rieke, Dave Spiegel, Scott Tremaine and an anonymous referee for useful conversations and comments.  This research has been supported in part by World Premier International Research Center Initiative, MEXT, Japan.

\bibliographystyle{natbib}

\end{document}